# On the fundamental conceptual problems in magnetism

Czeslaw Rudowicz[*]

*Department of Physics and Materials Science, City University of Hong Kong, 83 Tat Chee Avenue, Kowloon, Hong Kong SAR, People' s Republic of China*

**Abstract**

An extensive survey of misinterpretations and misconceptions concerning presentation of the hysteresis loop for ferromagnetic materials occurring in undergraduate textbooks as well as in recent magnetism literature has been carried out. This survey has also revealed a number of intricacies and conceptual problems concerning the fundamental aspect of magnetism. This note provides a critical analysis of these problems aimed at clarification of the intricacies and preventing further proliferation of the confusion in question.

*Address for correspondence: Tel.:+852-2788-7787; Fax:+852-2788-7830; E-mail address: apceslaw@cityu.edu.hk (C.Rudowicz).

## 1. Introduction

Recently an extensive work has been carried out to identify and clarify the misconceptions and misinterpretations concerning the magnetic hysteresis graphs appearing in college and university textbooks [1] as well as in recent magnetism-related research literature [2]. During this work it has turned out that a number of fundamental conceptual problems concerning magnetism have crept into the literature; an illustrative example is the Section "Fundamentals of magnetism" in the overview [3] of the magnetic systems used in biological high-gradient magnetic separation. These misconceptions do not bear on the validity of the technical aspects of magnetic hardware design intended for non-experts in magnetism [3]. Nevertheless in order to prevent further proliferation of the confusion in question, in this note we provide pertinent quotes and discuss the questionable statements in [3]. Such statements contribute to additional misconceptions or oversimplifications to those discussed in [1, 2]. Specific problems concerning the presentation of the magnetic hysteresis graphs in [3] have been dealt with in [2].

## 2. Origin of magnetism in insulators and metals

For the sake of the later discussion, it is worth to recall the following fundamental aspects. The microscopic origin of all magnetic properties in matter is due to the **quantum** phenomena. Electronic magnetism is due to the spin ($s_i$) and orbital ($l_i$) angular momenta of individual electrons and the associated magnetic moments. The latter may be *'localized'* on the paramagnetic ions in insulating crystals or *'itinerant'* as in the case of the conduction electrons in metals. Basically the nature of electronic magnetism in magnetic materials is twofold.

(a) In **insulators**, the paramagnetic ions with unpaired electrons exhibit permanent magnetic dipole moments due to the total spin momentum $S$ ($= \Sigma\, s_i$; the summation is over all electrons in a given unfilled electronic shell) or the total magnetic momentum $J$ ($= L + S$) of an ion. Here the different types of magnetic ordering arise due to specific alignment of the atomic magnetic moments (**m**) *'localized'* on the paramagnetic ions. The magnetization within insulating ferromagnets is produced by the ***internal* effective** field (molecular field $B_e$) acting on each magnetic atom and which is due to all other magnetic atoms. The molecular field $B_e$ concept was introduced by Weiss in 1907, i.e. long before the advent of quantum mechanics. Weiss assumed that $B_e$ is proportional to the magnetization M: $B_e = \lambda M$, where $\lambda$ is the Weiss molecular field constant. Later this concept found justification and explanation in terms of the Heisenberg exchange interaction Hamiltonian.

(b) In **metals**, the complex interaction between the nearly free electrons forming the Fermi sea and the mostly localized electrons from unfilled electronic shells (3d, 4f, or 5f) produces a collective magnetic state, e.g. band ferromagnetism or band antiferromagnetism. Here the 'magnetic ordering' has different meaning than in the case of insulators. It rather refers to redistribution of the conduction electrons with spin up and down in separate bands. Hence the magnetization in magnetic metals is mainly due to the excess of electrons, arranged in electronic bands, with a particular orientation of spins over the opposite orientation. Subtle interactions between the localized electrons and the conduction electrons, which are *'itinerant'*, i.e. not localized on particular ions, also play a role in magnetism of metals. Hence, the origin of magnetization and the magnetic interactions underlying its mechanism are different in insulators and metals. Regrettably the distinction between the insulating magnetic materials and magnetic metals is often blurred in textbooks to the point of confusion between the basic underlying concepts (for a review, see [1]).

## 3. Fundamental conceptual problems

*(1) Origin of magnetism*

*"In magnetic materials, a magnetic field is produced because of the movement of electrons within the material, which produces the field around the material and a magnetization effect within it."* [3]

The **classical** idea of electric current (i.e. '*movement of electrons'*) inducing magnetic field, which in turn produces magnetization, is borrowed from electromagnetism. However, it cannot explain the magnetization in ferromagnetic materials. The above statement creates an impression that '*the field around the material*' and '*a magnetization effect within it* (the material)' are two unrelated quantities, whereas in fact the former is a consequence of the latter, as explained in Section 2. Had the sequence been changed to '*which produces a magnetization effect within it and hence the field around the material"* no such confusion would arise.

*(2) Nature of atomic magnetic moments*

(i) *"With no external influence, atomic magnetic moments align themselves in their least-energy state and cancel each other internally so a material displays no net magnetic poles (and thus, no external magnetic field)."* [3]

(ii) *"Atomic magnetic moments are partially the result of electron orbits, so as temperature rises and the orbital path lengthens the strength of the magnetic moment decreases."* [3]

The first of the two consecutive sentences (i) is acceptable and pertains to paramagnetism of **insulators**, where the temperature randomizes the directions of the atomic magnetic moments localized on magnetic atoms. Hence the total magnetization due to the permanent dipole moments is zero. However, the second one (ii) is inappropriate from the quantum point of view, since the notion of *"the lengthening of the orbital path"* (supposedly with increasing temperature) cannot be justified by quantum mechanics. Moreover, the second sentence (ii) taken in the context of the first one (i) can hardly explain the nature of paramagnetism, neither of insulators nor of metals (see Section 2). The atomic magnetic moments can be visualized in the former case as arrows of the same length, but oriented in different directions. Hence, the explanation in [3] for the decrease of the magnetic moment with increasing temperature as due to the fact that *'the orbital path lengthens'* makes no sense.

*(3) Classification of magnetic materials*

*"Ferromagnetism, antiferromagnetism and ferrimagnetism are ordered states; diamagnetism and paramagnetism are transient states that exist as a result of an applied magnetic field."* [3]

Diamagnetism is the property of all matter, whereas paramagnetism is the property of some materials. Note that all ferromagnets, antiferromagnets, and ferrimagnets above certain transition temperature become paramagnetic. Hence diamagnetism and paramagnetism are not *"transient states"* but inherent properties, which exist irrespective the applied field is zero or non-zero. However, the magnetic susceptibility, by which we distinguish various kinds of magnetic properties, can be measured (i.e. "observed") by application of an applied magnetic field using a range of experimental techniques. It is the magnetization **M** induced by the applied magnetic field in diamagnets and paramagnets, which may be considered as *'transient'*, but not the diamagnetic and paramagnetic property itself.

*(4) Explanation of ferromagnetism*
   (i) *"The atomic moments align parallel to each other, because of an exchange interaction between electrons – there is a significant imbalance in the electron bands of coupled atoms, and there is therefore a large spontaneous magnetization $M_s$ present."* [3]
   (ii) *"This [$M_s$] is typically $10^4$ larger than the field [B] generated by the total magnetic moment [$\Sigma m_i$] of the ferromagnetic materials."* [3]

The origin of magnetism in insulators and that in metals discussed in Section 2 (and hence the associated two different mechanisms of magnetization) are confusingly mixed up in the above sentence (i). While the first part pertains to the nature of magnetization in insulators, the second parts attempts to explain the nature of magnetization in metals. Such statements only reinforce the confusion between the localized and itinerant sources of permanent magnetization. In the context of the first sentence (i) quoted above, the logic and the numerical value in the immediately following sentence (ii) are questionable. Denoting the physical quantities as shown by us above in brackets and using the relations (in SI units): $B = \mu_o M$ and $M = (\Sigma m_i)/V$ one obtains the relation: $M_s \sim 10^4 B = 10^4 \mu_o (\Sigma m_i)/V = 10^4 \mu_o M_s$, which indicates the confusion between the quantities involved.

*(5) Explanation of paramagnetism*
   (i) *"In the absence of an external magnetic field, the electron energy bands of a paramagnetic material are equally populated with spin 'up' and spin 'down' electrons."* [3]
   (ii) *"This paramagnetic effect has a temperature dependence, since the magnitude of the induced magnetic field* [actually, the **magnetization** - see below] *is limited by randomization due to thermal agitation within the atom."* [3]

The two sentences quoted above appear in the paragraph on paramagnetism in [3]. The simplified presentation without making a distinction between the nature of electronic magnetism in metals and insulators can only bring about more confusion between the concepts applicable to metals used in the sentence (i), i.e. energy bands and single electron spins and those applicable to insulators used in the sentence (ii), i.e. localized magnetic moments as evidenced by the phrase *within the atom*. Moreover, linking *'paramagnetic effect'* directly with *'the induced magnetic field'* amounts to missing the prime physical quantity, i.e. the induced magnetization **M**, which in turn yields the induced magnetic field intensity inside the sample: $\mathbf{B} = \mu_o \mathbf{M}$, in addition to the applied field intensity $\mathbf{B} = \mu_o \mathbf{H}$.

*(6) Distinction between 'external', 'internal', and 'applied' quantities and explanation of the ferromagnetic domains*

(i) *"Some, like iron, cobalt and nickel can exhibit strong <u>external</u> [a] magnetic field under certain conditions."* [3]

(ii) *"With no <u>external</u> [b] influence, atomic magnetic moments align themselves in their least-energy state and cancel each other internally so a material displays no net magnetic poles (and thus, no <u>external</u> [a] magnetic field)."* [3]

The usage of the term '*external*' in two different contexts denoted by subscripts [a] and [b] above may be noted in the above sentences. In fact, there is an *internal* magnetization **M**, which is responsible for the observed magnetic field inside the sample: **B** = $\mu_o$(**H** + **M**). In the other example (ii), the two meanings are mixed up in the same sentence. While in the first instance '*external*' [b] refers to a truly 'external' quantity, in the second instance '*external*' [a] refers actually to an **internal** quantity, i.e. the internal magnetization **M**, which is responsible for the observed magnetic field *inside* and *outside* the sample in the absence of the applied magnetic field. In order to avoid confusion the term '*external magnetic field*' should be rather reserved for the '*applied field*', which is truly **external** (i.e. with respect to the sample). Otherwise the internal properties of materials may be confused with the external factors acting on the materials. The second sentence (ii) taken in the context of the first one (i) refers obviously to the existence of **the ferromagnetic domains.** Such an indirect way of presentation of an important physical concept is not meaningful in general, although it may be due to the brevity of the presentation in [3].

## 4. Conclusions

Several fundamental conceptual problems concerning magnetism presented in [3] have been analyzed. An appropriate brief presentation for each concept has been given here in order to clarify the misinterpretations and misconceptions. It is hoped that the present critical analysis of the fundamental conceptual problems together with the clarification of the misconceptions and misinterpretations concerning the magnetic hysteresis graphs appearing in college and university textbooks [1] as well as in recent magnetism-related research literature [2] will help preventing further proliferation of the confusion in question.

**Acknowledgements**

This work was supported by the City University of Hong Kong through the research grant: QEF # 8710126. Technical assistance from Miss H.W.F. Sung is gratefully acknowledged. Comments from Dr Hatch, which have helped to improve presentation of some issues, are also appreciated.